\documentclass[sigconf, screen, dvipsnames, authorversion]{acmart}

\usepackage[noend]{algorithmic}
\usepackage{algorithm}
\usepackage{multirow}
\usepackage{amsmath}
\usepackage{bbm}
\usepackage{booktabs}
\usepackage[inline]{enumitem} 
\usepackage{subcaption}
\usepackage{bm} 
\usepackage{xcolor}

\usepackage{hyperref}
\usepackage{lipsum}

\usepackage{wrapfig}

\usepackage{arydshln}
\makeatletter
\def\adl@drawiv#1#2#3{
        \hskip.5\tabcolsep
        \xleaders#3{#2.5\@tempdimb #1{1}#2.5\@tempdimb}%
                #2\z@ plus1fil minus1fil\relax
        \hskip.5\tabcolsep}
\newcommand{\cdashlinelr}[1]{%
  \noalign{\vskip\aboverulesep
          \global\let\@dashdrawstore\adl@draw
          \global\let\ adl@draw\adl@drawiv}
  \cdashline{#1}
  \noalign{\global\let\adl@draw\@dashdrawstore
          \vskip\belowrulesep}}
\makeatother

\usepackage{tikz}
\usetikzlibrary{automata, arrows, bayesnet, bending}

\usepackage{tikzscale}

\DeclareMathOperator*{\argmax}{arg\,max}

\copyrightyear{2024}
\acmYear{2024}
\setcopyright{acmlicensed}\acmConference[RecSys '24]{18th ACM Conference on Recommender Systems}{October 14--18, 2024}{Bari, Italy}
\acmBooktitle{18th ACM Conference on Recommender Systems (RecSys '24), October 14--18, 2024, Bari, Italy}
\acmDOI{10.1145/3640457.3688132}
\acmISBN{979-8-4007-0505-2/24/10}

\begin{document}

\title[{Multi-Objective Recommendation via Multivariate Policy Learning}]{Multi-Objective Recommendation~\\ via Multivariate Policy Learning}

\author{Olivier Jeunen}
\affiliation{
  \institution{ShareChat}
  \country{United Kingdom}
}
\author{Jatin Mandav}
\affiliation{
  \institution{ShareChat}
  \country{India}
}
\author{Ivan Potapov}
\affiliation{
  \institution{ShareChat}
  \country{United Kingdom}
}
\author{Nakul Agarwal}
\affiliation{
  \institution{ShareChat}
  \country{India}
}
\author{Sourabh Vaid}
\affiliation{
  \institution{ShareChat}
  \country{India}
}
\author{Wenzhe Shi}
\affiliation{
  \institution{ShareChat}
  \country{United Kingdom}
}
\author{Aleksei Ustimenko}
\affiliation{
  \institution{ShareChat}
  \country{United Kingdom}
}

\begin{abstract}
Real-world recommender systems often need to balance multiple objectives when deciding which recommendations to present to users.
These include behavioural signals (e.g. clicks, shares, dwell time), as well as broader objectives (e.g. diversity, fairness).
Scalarisation methods are commonly used to handle this balancing task, where a weighted average of per-objective reward signals determines the final score used for ranking.
Naturally, \emph{how} these weights are computed exactly, is key to success for any online platform.

We frame this as a decision-making task, where the scalarisation weights are \emph{actions} taken to maximise an overall North Star reward (e.g. long-term user retention or growth).
We extend existing policy learning methods to the continuous multivariate action domain, proposing to maximise a pessimistic lower bound on the North Star reward that the learnt policy will yield.
Typical lower bounds based on normal approximations suffer from insufficient coverage, and we propose an efficient and effective policy-dependent correction for this.
We provide guidance to design stochastic data collection policies, as well as highly sensitive reward signals.
Empirical observations from simulations, offline and online experiments highlight the efficacy of our deployed approach.
\end{abstract}

\maketitle

\section{Introduction \& Motivation}
Recommender systems are crucial tools that empower online platforms to connect users to content they enjoy---be it on music and video streaming platforms, e-commerce websites, social media applications, or others.
Practical implementations of such systems serve broad and diverse use-cases, with a few common tendencies.
One of those, is that recommendations are rarely centred around a \emph{single} objective.
Indeed, streaming platforms might want to optimise both short-term and long-term engagement whilst ultimately targeting retention and lifetime customer value~\cite{Briand2024,Mehrotra2020,Mehrotra2018,Bugliarello2022,Tang2023}; e-commerce platforms need to balance clicks, add-to-carts and conversions with possible returns and advertising income~\cite{Gu2020,Lin2019,Wan2018}; and social media platforms encounter similar challenges~\cite{Sagtani2023,Sagtani2024}.

Another recent tendency is that such systems are increasingly often framed as \emph{decision-making} instead of \emph{prediction} systems.
Indeed, predictions about user-item affinities typically serve to inform real-time decisions about recommendations, rankings, advertisements, and so forth.
The decision-making lens allows us to reason about \emph{consequences} of such systems, and has been gaining popularity in recent years~\cite{CONSEQUENCES2022}.
In the context of real-world platforms, it allows us to frame changes in key online metrics as the consequences of recommendation decisions, which in turn allows us to optimise those online metrics directly~\cite{Jeunen2021_Thesis, Jeunen2024_nDCG}.
Approaches that leverage the decision-making literature and frame recommendation as a \emph{bandit} learning task have led to several practical successes~\cite{McInerney2018,chen2019top,chen2021, chen2022actorcritic, Dong2020, ma2020off, Mehrotra2020,Bendada2020, Su2024, Briand2024}.
Most common is the \emph{off-policy} or \emph{counterfactual} family of approaches, as they allow practitioners to learn and evaluate models \emph{offline} before \emph{online} deployment~\cite{vandenAkker2024}.
Such methods typically leverage some form of algorithmic \emph{pessimism}, optimising a lower bound on the reward they will yield~\cite{Jeunen2021_Pessimism,Jeunen2023,Swaminathan2015_BLBF}.

The majority of the (off-policy) bandit literature focuses on \emph{single} rewards, but \emph{multi}-objective bandit approaches have been proposed as well~\cite{Drugan2013,Busa-Fekete2017}.
In the context of multi-objective recommendation with bandits, a few specialised solutions have been proposed recently.
These either focus on artist objectives in music streaming platforms~\cite{Mehrotra2020}, exposure fairness objectives in top-$K$ settings~\cite{Jeunen2021_TopK}, or long-term value via reinforcement learning~\cite{Zhang2022}.

In this work, we propose a general multi-objective recommendation approach that frames the optimisation of the \emph{weights} that are given to individual objectives as a decision-making problem.
To this end, we leverage the Counterfactual Risk Minimisation (CRM) principle to optimise a lower bound on the reward a recommendation policy will yield, thereby extending it to a multivariate continuous action domain~\cite{Swaminathan2015_BLBF}.
We show that existing approaches to construct the lower bound, based on the Central Limit Theorem (CLT), provide insufficient coverage in finite sample scenarios, and propose an efficient and effective policy-dependent correction for this problem.
In doing so, we combine existing elements in the literature to derive a novel estimator for the Effective Sample Size~\cite{Martino2017,Owen2013}.

Reproducible empirical observations on synthetic data show that our proposed approach reduces the required sample size for the confidence interval to have sufficient coverage by a factor of up to 60, significantly reducing the cost of randomisation that is required for off-policy learning to work effectively.
We discuss practical considerations for designing logging policies in multivariate continuous domains, showing that the ``\emph{curse of dimensionality}'' makes uniform randomisation less useful than practitioners would typically assume.
Furthermore, we show how ``\emph{learnt metrics}'' give rise to highly sensitive reward signals that further improve effectiveness by improving statistical power~\cite{Jeunen2024_Learning}.
Using real-world data from a large-scale short-video platform, we perform offline experiments that highlight the promise of our learnt policies.

Online experiments on two platforms with monthly user-bases over 160 million each, show that our approach significantly outperforms existing alternatives on multiple key metrics, bringing significant value to the business. 
The alignment between our off- and online experimental results underscores the value of the bandit learning paradigm, and the decision-making lens in general.
\section{Multi-Objective Recommendation}
Users interact with content on the web through various modalities.
Online content marketplaces such as Instagram, Reddit, TikTok or ShareChat allow users to view, click, (dis)like, save, share and comment on items that are presented to them by the recommender system.
Such systems are typically not optimised for a \emph{single} type of interaction, but rather aim to maximise a set of positive behaviours through multi-objective optimisation techniques.
Decidedly the most common approach in practice is scalarisation~\cite{Hwang1979}, where the multi-objective optimisation problem is recast with a single scalar objective.
Different parameterisations for the function that maps the multiple objectives to a single scalar, then give rise to different Pareto-optimal solutions.
Pareto-optimality in multi-objective recommendation applications is a well-studied topic~\cite{Ribeiro2012,Lin2019,Xie2021}.
When the Pareto-front is convex, a \textit{linear} scalarisation function is sufficient to represent \emph{all} Pareto-optimal solutions.
Naturally, this does not tell us \emph{how} the weights of the linear function should be chosen, or where on the Pareto-front the business should position itself to optimise long-term business goals.

Formally, assume we need to decide on a recommendation $i\in\mathcal{I}$ to show to a user $u\in\mathcal{U}$, in an attempt to optimise $d\in \mathbb{N}$ objectives represented as $f(u,i)$.
A multi-objective decision-making procedure with linear scalarisation weights $a \in \mathbb{R}^{d}$ can then be described as finding the items that satisfy:
\begin{equation}\label{eq:scalarisation}
    \max_{i\in \mathcal{I}} \sum_{k=1}^{d} a_{k}\cdot f_{k}(u,i).
\end{equation}
Naturally, in top-$n$ settings we can replace the $\max$-operation with a ${\tt sort}$-operation.
This type of procedure is exceedingly common in deployed recommender systems on the web.
Indeed: public accounts describe Facebook and TikTok's systems in this way~\cite{Smith2021,Merrill2021} and Twitter even open-sourced their exact weights~\cite{Twitter2023}.
\citeauthor{Mehrotra2020}~\cite{Mehrotra2020} describe how Spotify leverages the Generalised Gini Index~\cite{Busa-Fekete2017} aggregation function to decide on the weights, motivated as preserving fairness between user- and artist-centric objectives.
\citeauthor{Zhang2022} describe a reinforcement learning approach for Tencent that aims to find weights that maximise long-term user satisfaction~\cite{Zhang2022}.
\citeauthor{Milli2023} study strategic behaviour of users and content providers that might result from the weights being chosen~\cite{Milli2023}.
\citeauthor{Jannach2023} present a taxonomy of \emph{types} of objectives that might be considered in multi-objective recommendation settings, whilst outlining open challenges~\cite{Jannach2023}.

Our work complements the existing literature by describing a policy learning approach to align the scalarisation weights with an over-arching North Star reward, such as long-term growth or revenue.
Crucially, this can help guide online platforms to decide \emph{where} on the Pareto-front they wish to position themselves.
Additional to technical contributions improving the robustness of existing policy learning methods, we present extensive experimental results that highlight its real-world effectiveness.

\section{Multivariate Off-Policy Learning}\label{sec:crm}
We want to learn a $d$-dimensional weight vector $a \in \mathbb{R}^{d}$, which we will refer to as an action $A$, in line with the bandit literature~\cite{Lattimore2020}.
The space of all possible actions, i.e. $\mathbb{R}^{d}$, is then denoted by the calligraphic $\mathcal{A}$.
When taken, an action yields a reward $R$ (e.g. long-term retention, growth, revenue).
The goal is to find the weights that maximise rewards.
Specifically, we frame this as a policy learning problem, where a policy $\pi_{\theta}$ describes a distribution over actions $\pi_{\theta}(A) \equiv\mathsf{P}(A|\pi,\theta)$.
Note that the probabilistic view is general, but $\pi_{\theta}$ is allowed to be deterministic.
We wish to learn the parameters that maximise the expected reward under the learnt policy:
\begin{equation}\label{eq:goal}
\theta^{\star} = \argmax_{\theta \in \Theta}\mathbb{E}_{a \sim \pi_{\theta}}[R].\end{equation}

This is often attained via gradient ascent on an importance sampling or Inverse Propensity Score (IPS) weighting estimator, where we leverage samples from a given \textit{data collection} or ``\textit{logging}'' policy to allow for unbiased counterfactual estimation:
\begin{equation}\label{eq:IPS}
    \mathbb{E}_{a \sim \pi_{\theta}} [R] = \mathbb{E}_{a \sim \pi_{0}} \left[ \frac{\pi_{\theta}(a)}{\pi_{0}(a)}R\right].
\end{equation}
Being able to estimate the left-hand side of Eq.~\ref{eq:IPS} without actually needing to deploy $\pi_{\theta}$, is what makes Off-Policy Learning (OPL) attractive and desirable.
We only need a single assumption to make this work: the logging policy $\pi_{0}$ should have common support with the target policy $\pi$. I.e. $\forall a \in \mathcal{A}:\pi(a)>0 \implies \pi_{0}(a) > 0$.

Note that this implies that the \emph{logging} policy \emph{should} be stochastic.
Whilst a natural consideration for OPL researchers and practitioners, the choice of logging policy is crucial.
We discuss concerns when constructing data collection policies in Section~\ref{sec:logging}.

Suppose we have a logged dataset $\mathcal{D} = \{(a_{i},r_{i})_{i=1}^{N}\}$, where actions were sampled according to the logging policy: $a_{i}\sim\pi_0(A)$.
Then, we can get a sample estimate for the quantity in Eq.~\ref{eq:IPS} as:
\begin{equation}\label{eq:IPS-estimate}
    \widehat{V}_{\rm IPS}(\pi_{\theta},\mathcal{D}) = \frac{1}{|\mathcal{D}|} \sum_{(a,r) \in \mathcal{D}} \frac{\pi_{\theta}(a)}{\pi_{0}(a)}r.
\end{equation}

Eq.~\ref{eq:IPS} gives rise to an unbiased estimator for the optimisation objective in Eq.~\ref{eq:goal}, formulated in Eq.~\ref{eq:IPS-estimate}.
Its variance, however, is often problematic.
A common solution is to introduce a multiplicative control variate to the IPS estimator, and perform \emph{self-normalised} importance sampling instead:
\begin{equation}\label{eq:SNIPS}
    \mathbb{E}_{a \sim \pi_{\theta}} [R] =  \frac{\mathbb{E}_{a \sim \pi_{0}} \left[ \frac{\pi_{\theta}(a)}{\pi_{0}(a)}R\right]}{\mathbb{E}_{a \sim \pi_{0}} \left[ \frac{\pi_{\theta}(a)}{\pi_{0}(a)}\right]}.
\end{equation}
Indeed, the denominator on the right-hand side of this equation equals 1 if the common support assumption is met~\cite{London2022}.
Sample average approximations for this Self-Normalised IPS (SNIPS) estimator typically enjoy significantly reduced variance over the IPS estimator, whilst remaining asymptotically unbiased~\cite{Kong1992}: 
\begin{equation}
    \widehat{V}_{\rm SNIPS}(\pi_{\theta},\mathcal{D}) = \frac{\sum_{(a,r) \in \mathcal{D}} \frac{\pi_{\theta}(a)}{\pi_{0}(a)}r}{\sum_{(a,r) \in \mathcal{D}} \frac{\pi_{\theta}(a)}{\pi_{0}(a)}}.
\end{equation}
Reducing the variance of the estimator does not solve all problems.
Indeed, we run into Goodhart's law, often paraphrased as: ``\textit{when a measure becomes a target, it ceases to be a good measure}''~\cite{Goodhart1984}.
In other words, maximising it directly likely leads to overfitting.

To deal with this, the Counterfactual Risk Minimisation (CRM) paradigm proposes to optimise a lower bound on the true reward instead, either for IPS~\cite{Swaminathan2015_BLBF} or SNIPS~\cite{Swaminathan2015}:
\begin{equation}\label{eq:crm}
\theta^{\star} = \argmax_{\theta \in \Theta} \widehat{V}_{\rm (SN)IPS}(\pi_{\theta},\mathcal{D}) - \lambda{\rm Var}\left(\widehat{V}_{\rm (SN)IPS}(\pi_{\theta},\mathcal{D})\right).\end{equation}

The variance penalisation term is estimated empirically for IPS~\cite{Maurer2009}, and approximated using the delta method for SNIPS~\cite[Eq. 9.9]{Owen2013}.
Naturally, these variance estimates are only useful when confidence intervals constructed using them have the required coverage.
Whilst the Central Limit Theorem (CLT) ensures this \emph{eventually}, it might fall short in finite sample scenarios.
We discuss this in detail and propose effective extensions in Section~\ref{sec:ESS}.

Another important aspect to consider is that our action space $\mathcal{A}$ is multivariate and continuous.
This entails that we perform importance sampling with probability \emph{density} functions.
In the case of deterministic distributions, these are Dirac delta functions.
In other words, the probability that an observed action $a \sim \pi_{0}$ has non-zero density under a \emph{deterministic} learnt policy $\pi_{\theta}$ \emph{exactly} is practically non-existent.
Kernel smoothing techniques have been proposed to deal with this~\cite{Kallus2018}.
A natural choice for the kernel is the Gaussian density function. Letting $\theta\equiv\mu$ and treating the kernel bandwidth $\Sigma$ as a matrix hyper-parameter, this is given by:
\begin{equation}\label{eq:gausskernel}
\pi_{\theta}(a) \approx \mathsf{P}(A=a|\mathcal{N}(\mu,\Sigma)) = \frac{\exp \left( -\frac{1}{2} (a-\mu)^\top \Sigma^{-1} (a- \mu) \right)}{(2\pi)^{\frac{d}{2}} \sqrt{|\Sigma|}}.
\end{equation}
For $\lim_{\Sigma \to \bm{0}}$, the kernel converges to the deterministic distribution, and hence the estimate is unbiased.
Nevertheless, this can lead to excessively high variance because of a low \emph{effective} sample size~\cite[\S 9.3]{Owen2013}.
For $\lim_{\sigma \to \infty\bm{I}}$, the kernel approximates a uniform distribution. Whilst this minimises variance, it significantly increases bias.
The Gaussian kernel permits an intuitive understanding: we estimate \emph{as if} we would deploy a Gaussian policy, when in reality we ignore its covariance matrix and simply deploy its mean deterministically.
This gives rise to an intuitive way to visualise estimates of policy value: we vary values of $\Sigma$ and consider their trend across the bias--variance trade-off.
We introduce, discuss, and visualise such results for off-policy evaluation in Section~\ref{sec:offline}.

Finally, we note that the Gaussian kernel is one of many alternatives to consider, albeit an intuitive option.
Any multivariate continuous probability distribution gives rise to a density function that can be used either directly to optimise a stochastic policy, or as a kernel smoothing function when optimising a deterministic policy.
The problems we describe in Section~\ref{sec:logging} that arise from uniform logging distributions, hold for all distributions and kernels that have an unrestricted domain.

\section{Improving the CRM Lower Bound}\label{sec:ESS}
We require accurate variance estimates for two reasons: (1) during evaluation, we want to ensure statistical significance of our results, obtaining confidence intervals that exhibit the expected coverage levels so we can make meaningful statements about probabilities, and (2) during learning, we make use of sample variance penalisation schemes that assume accurate variance estimates to provide sensible lower bounds. As such: we require an estimator for ${\rm Var}\left(\widehat{V}_{\rm (SN)IPS}(\pi_{\theta},\mathcal{D})\right)$ that implies confidence intervals with expected coverage.
That is, when $\lambda\approx1.96$ in Eq.~\ref{eq:crm}, the lower bound holds in 95\% of cases.
Traditionally used variance estimators are based on Gaussian approximations, motivated through the CLT.
For $\widehat{V}_{\rm IPS}$, the sample variance with $N$ as the sample size is given by:
\begin{equation}\label{eq:ips_var}
    \widehat{{\rm Var}}\left(\widehat{V}_{\rm IPS}(\pi_{\theta},\mathcal{D})\right) = \frac{\sum_{(a,r)\in \mathcal{D}}\left(\frac{\pi_{\theta}(a)}{\pi_{0}(a)}r - \widehat{V}_{\rm IPS}(\pi_{\theta},\mathcal{D})\right)^{2}}{{N-1}} .
\end{equation}

Because the $\widehat{V}_{\rm SNIPS}$ estimator deals with a \emph{ratio} of two expectations, we need to resort to the delta method to approximate its variance~\cite[Eq. 9.9]{Owen2013}:
\begin{equation}\label{eq:snips_var}
    \widehat{{\rm Var}}\left(\widehat{V}_{\rm SNIPS}(\pi_{\theta},\mathcal{D})\right) = \frac{\sum_{(a,r)\in \mathcal{D}}\left(\frac{\pi_{\theta}(a)}{\pi_{0}(a)}r - \frac{\pi_{\theta}(a)}{\pi_{0}(a)}\widehat{V}_{\rm SNIPS}(\pi_{\theta},\mathcal{D}) \right)^{2}}{(N-1)\left(\frac{1}{|\mathcal{D}|}\sum_{(a,r)\in \mathcal{D}} \frac{\pi_{\theta}(a)}{\pi_{0}(a)} \right)^{2} } .
\end{equation}

\subsection{Effective Sample Size Corrections}
Whilst the CLT justifies the use of such estimates to obtain confidence intervals for $\widehat{V}_{\rm (SN)IPS}$ in large-sample scenarios, it does not tell us when the sample size is sufficient.
As \citeauthor{Bottou2013} write: ``\emph{central limit convergence might occur
too slowly to justify such confidence intervals}''~\cite{Bottou2013}.
We empirically observe this phenomenon, and highlight it in our experiments.
\citeauthor{Bottou2013}'s proposed solution is two-fold~\cite{Bottou2013}: reduce the variance of $\widehat{V}_{\rm IPS}$ by clipping the importance weights~\cite{Ionides2008,Gilotte2018}, bound the bias this introduces to the estimator, and provide an \emph{inner} and \emph{outer} confidence interval.
Whilst this can be effective for \emph{evaluation}, we wish to obtain confidence intervals that we can plug into Eq.~\ref{eq:crm} to obtain a lower bound with improved empirical coverage.  
As such, the alternative estimator we wish to construct has two desiderata:
\begin{enumerate*}
    \item improved coverage at small sample sizes, and
    \item convergence to the Gaussian estimator at large sample sizes (i.e. \emph{consistency} with the CLT).
\end{enumerate*}
Indeed, inflating the variance estimate will help (1), but leave much to be desired for (2).

To satisfy property (1), we leverage the Effective Sample Size (ESS)~\cite[\S 9.3]{Owen2013}.
The ESS estimates the number of independent samples from the target policy that would be equally informative as the actual weighted samples from the logging policy that we have~\cite{Elvira2022}. 
When the ESS is low, we can expect the variance estimates in Eqs.~\ref{eq:ips_var}--\ref{eq:snips_var} to be \emph{under}-estimates.
When the ESS is high (i.e. maximally $N$ when $\pi_{\theta}\equiv\pi_{0}$), we are not losing any statistical efficiency due to importance sampling, and we can only hope for CLT convergence.
To satisfy property (2), we formulate our estimator as a multiplicative correction factor to the original sample size.
Thus, as the sample size grows, we retain consistency and coverage guarantees at larger sample sizes (from the CLT), whilst exhibiting improved coverage at smaller sample sizes (which we show empirically in Section~\ref{sec:synthetic}).
We define the ESS-corrected sample size as:
\begin{equation}
    \widetilde{N}= 1 + \frac{N(\widehat{\rm ESS}-1)}{\widehat{\rm ESS}}.
\end{equation}
From this formula, it is clear that $\widetilde{N}$ is monotonically increasing as a function of $N$, and that $\widetilde{N}\equiv N$ when $\widehat{\rm ESS}\equiv N$.
Furthermore, when ${\rm ESS}>1$, as $N$ grows, Eq. 11 converges towards $N$. 
As a result, $\widetilde{N}$ is a consistent estimator for $N$, and as a consequence, variance estimates using $\widetilde{N}$ instead of $N$ are also consistent.

In Eqs.~\ref{eq:ips_var}--\ref{eq:snips_var}, we then simply replace $N$ with $\widetilde{N}$ and denote this as $\widehat{{\rm Var_{ESS}}}$.
This gives rise to a symmetric confidence interval as:
\begin{equation}\label{eq:better_LB}
   \widehat{V}_{\rm (SN)IPS}(\pi_{\theta},\mathcal{D}) \pm \Phi^{-1}\left(1-\frac{\alpha}{2}\right)\sqrt{\frac{\widehat{\rm Var_{ESS}}\left(\widehat{V}_{\rm (SN)IPS}(\pi_{\theta},\mathcal{D})\right)}{\widetilde{N}}}.
\end{equation}
Naturally, $\Phi^{-1}$ represents the quantile function of a standard normal distribution, such that the multiplicative factor roughly equals $1.96$ when $\alpha=0.05$ to obtain standard 95\% confidence intervals.

This yields a robust lower bound on the value of the policy that we wish to maximise.
We expect the ESS correction to appropriately inflate the width of the confidence interval when the ESS is low, and the CLT is thus unlikely to hold, improving the estimators' empirical coverage in low sample-size scenarios.

\subsection{Estimating the Effective Sample Size}
A widespread estimator for the ESS is given by~\cite[Eq. 9.13]{Owen2013}:
\begin{equation}
    P^{2}_{N} = \frac{1}{\sum_{i}\overline{w}^{2}_{i}}, \text{~where~}\overline{w}_{i} = \frac{w_{i}}{\sum\limits_{(a_{j},r_{j}) \in \mathcal{D}} w_{j}} = \frac{\frac{\pi_{\theta}(a_i)}{\pi_0(a_i)}}{\sum\limits_{(a_{j},r_{j}) \in \mathcal{D}} \frac{\pi_{\theta}(a_j)}{\pi_0(a_j)}}.
\end{equation}
Whilst common and intuitive, this estimator has flaws.
First, \citeauthor{Owen2013} remarks that the effectiveness of importance sampling depends on the distribution of $R$, which is not captured here.
They propose to define a reward-specific estimator as~\cite[Eq. 9.17]{Owen2013}:
\begin{gather}
    P^{R-2}_{N} = \frac{1}{\sum_{i}\widetilde{w}^{2}_{i}},\\
    \text{~where~} \widetilde{w}_{i} = \frac{w_{i}|r_{i}|}{\sum\limits_{(a_{j},r_{j}) \in \mathcal{D}} w_{j}|r_{j}|} = \frac{\frac{\pi(a_i)}{\pi_0(a_i)}|r_{i}|}{\sum\limits_{(a_{j},r_{j}) \in \mathcal{D}} \frac{\pi(a_j)}{\pi_0(a_j)}|r_{j}|}. \nonumber 
\end{gather}
\citeauthor{Martino2017} independently remark that $P^{2}_{N}$ is related to the Euclidean distance between the distribution of the normalised importance weights and uniform weights (i.e. when $\pi_{\theta}\equiv\pi_{0}$)~\cite{Martino2017}.
They derive alternative estimators arising from alternative discrepancy measures and evaluate them on a synthetic task, showing promising results for a formulation that considers the $\ell_{\infty}$-norm instead of the $\ell_{2}$-norm implied by the Euclidean distance measure (shown in Eq.~\ref{eq:dinf}).
We can equivalently extend this to obtain a reward-dependent estimator in Eq.~\ref{eq:rdinf}:\\
\noindent\begin{minipage}{0.5\linewidth}
    \begin{equation}\label{eq:dinf}
      D^{\infty}_{N} = \frac{1}{\max_{i}\overline{w}_{i}}, 
    \end{equation}
\end{minipage}%
\begin{minipage}{0.5\linewidth}
    \begin{equation}\label{eq:rdinf}
       D^{R-\infty}_{N} = \frac{1}{\max_{i}\widetilde{w}_{i}}.
    \end{equation}
\end{minipage}
Note that the latter provides a novel estimator for the ESS, combining existing elements in the literature~\cite{Owen2013,Martino2017}.
We can use any of these estimators to compute $\widehat{\rm ESS}, \widetilde{N}$ and the resulting confidence interval in Eq.~\ref{eq:better_LB}.
In Section~\ref{sec:synthetic}, we empirically verify C.I. coverage for all mentioned ESS estimators.

\section{Considerations for logging policies}\label{sec:logging}
Assuming stationarity, the optimal scalarisation weights will be deterministic.
That is, if we do not consider the explore--exploit trade-off that arises when learning over time, the density of the optimal policy $\pi_{\theta^{\star}}$ will be a Dirac-delta centred at a single point.
It comes natural, however, to assume that $\pi_{\theta^{\star}}$ is not static and the optimal weights might drift over time.
As discussed in Section~\ref{sec:crm}, the \emph{logging} policy that is responsible for collecting the samples we use to estimate Eq.~\ref{eq:better_LB} needs to be stochastic.
Thus, practitioners need to deploy a randomisation strategy that collects informative samples to allow for counterfactual estimation.
For general applications, \emph{uniform} distributions are attractive.
Indeed, they are intuitive and easy to implement, and the constant logging propensities $\pi_{0}(a)$ they imply significantly simplify both the modelling and engineering efforts required to run off-policy learning in production environments~\cite{vandenAkker2024}.
As such, they are commonly used.

Notwithstanding this, uniform distributions bring significant downsides as well.
In this Section, we wish to dissuade researchers and practitioners from using uniform logging policies in the multivariate continuous case, focusing on three arguments.

\subsection{Violating the common support assumption}
Suppose we have a vector of production weights $\bm{a_{0}} \in \mathbb{R}^{d}$, around which we wish to design a logging policy.
Suppose we wish to explore a relative $\epsilon\%$ range on either side of the current weights: hence we have $\pi_{0} \equiv \mathcal{U}\left((1-\frac{\epsilon}{100})\bm{a_{0}}; (1+\frac{\epsilon}{100})\bm{a_{0}}\right)$.
Whilst a very natural choice, it should be noted that the logging policy now has a restricted domain.
One might assume that if we are merely learning a deterministic target policy, this is not hugely problematic.
Nevertheless, even when we learn deterministic policies, we need to resort to kernel smoothing techniques such as the Gaussian kernel introduced in Eq.~\ref{eq:gausskernel}.
The Gaussian kernel implies a domain over $\mathcal{A}\equiv\mathbb{R}^{d}$, and hence, the common support assumption is violated:
\begin{equation}
 \forall a \in \mathbb{R}^{d}_{\setminus \left[(1-\frac{\epsilon}{100})\bm{a_{0}}; (1+\frac{\epsilon}{100})\bm{a_{0}}\right]}: \pi_{0}(a)=0\land\pi_{\theta}(a)>0. \nonumber    
\end{equation}
The main problem that this violation implies, is that the expected importance weight no longer equals 1~\cite{London2022}. Aside from the fact that this implies that the conventional IPS estimator is now biased, it brings along additional problems.
Indeed, baseline corrections now have a significant effect on the estimator, and should be avoided.
Usually, practitioners recentre observed rewards as a simple variance-reduction technique~\cite{Gupta2024_baselines}.
When the common support assumption holds, we can straightforwardly show that any constant scalar translation $\beta$ preserves the unbiasedness of the estimator:
\begin{gather} \beta + \mathop{\mathbb{E}}\limits_{a \sim \pi_{0}} \left[ \frac{\pi_{\theta}(a)}{\pi_{0}(a)}(R-\beta)\right] = \beta + \mathop{\mathbb{E}}\limits_{a \sim \pi_{0}} \left[ \frac{\pi_{\theta}(a)}{\pi_{0}(a)}R\right] - \mathop{\mathbb{E}}\limits_{a \sim \pi_{0}} \left[ \frac{\pi_{\theta}(a)}{\pi_{0}(a)}\beta\right] \nonumber\\ 
= \beta + \mathop{\mathbb{E}}\limits_{a \sim \pi_{0}} \left[ \frac{\pi_{\theta}(a)}{\pi_{0}(a)}R\right] - \beta\underbrace{\mathop{\mathbb{E}}\limits_{a \sim \pi_{0}} \left[ \frac{\pi_{\theta}(a)}{\pi_{0}(a)}\right]}_{\textcolor{Maroon}{\nequiv1}}\textcolor{Maroon}{\neq}\mathop{\mathbb{E}}\limits_{a \sim \pi_{0}} \left[ \frac{\pi_{\theta}(a)}{\pi_{0}(a)}R\right].\end{gather}
When we leverage a uniform logging policy with any kernel that has a domain that can extend beyond that of $\pi_{0}$, the common support assumption does not hold.
As a result, the equivalence of baseline corrections is now violated, and they should be avoided.

\subsection{Self-normalisation becomes unstable}
The SNIPS estimator introduced in Eq.~\ref{eq:SNIPS} also leverages the fact that $\mathbb{E}_{a \sim \pi_{0}} \left[ \frac{\pi_{\theta}(a)}{\pi_{0}(a)}\right] = 1.$
Because this equivalence no longer holds, the SNIPS estimator is no longer asymptotically unbiased.
More problematically, when leveraging this estimator for \emph{learning} purposes, we are incentivised to find a policy that minimises $\mathbb{E}_{a \sim \pi_{0}} \left[ \frac{\pi_{\theta}(a)}{\pi_{0}(a)}\right]$ (as it appears in the denominator of Eqs.~\ref{eq:SNIPS} and~\ref{eq:better_LB}).
This quantity can become arbitrarily small by moving $\pi_{\theta}$ to or beyond the borders of the logging policy’s domain.
Indeed: this \emph{decreases} $\mathbb{E}_{a \sim \pi_{0}} \left[ \frac{\pi_{\theta}(a)}{\pi_{0}(a)}\right]$ and \emph{increases} $\widehat{V}_{\rm SNIPS}$ as a result.
It should nevertheless intuitively be clear that this is undesirable behaviour: placing probability density \emph{away} from the observed data should not increase our reward estimate. 
This is closely related to the ``propensity overfitting'' phenomenon described by \citeauthor{Swaminathan2015}~\cite{Swaminathan2015}, which the uniform logging policy cannot avoid here.
As such, we cannot enjoy the variance reduction that self-normalisation brings without introducing a statistical bias that is problematic.

\begin{figure*}[t]
    \centering
    \vspace{-3ex}
    \includegraphics[width=\linewidth]{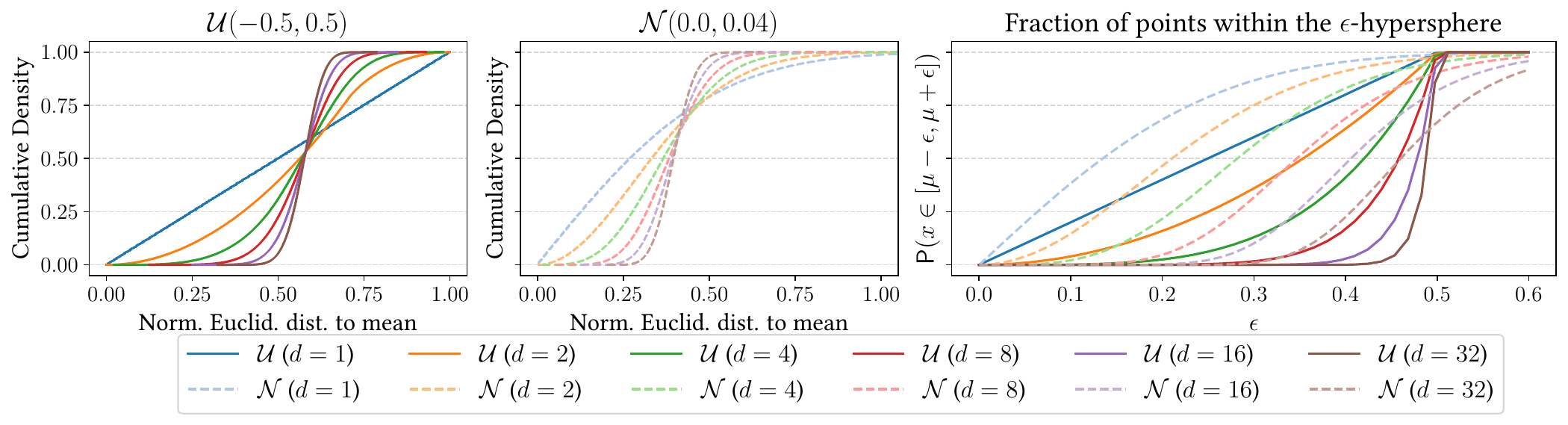}
    \caption{Visualising the curse of dimensionality and counterintuitive properties of the Uniform distribution in high dimensions. \textit{Left}: Whilst all distances are equally likely in a single dimension, this property quickly disappears in higher dimensions. \textit{Middle}: The Normal distribution exhibits similar but less pronounced behaviour, with a lower slope on the cumulative density. \textit{Right}: Uniform sampling disfavours a hypersphere around the mean. Normal sampling can help to partially alleviate this.}
    \label{fig:cod}
\end{figure*}
\subsection{The curse of dimensionality}
Finally, whilst the uniform distribution has certain intuitive and desirable properties in a \emph{single} dimension, they can become counterintuitive and disappear when we consider \emph{multiple} dimensions.
A key desirable property in the univariate case is that sampled points’ distances to the centre of the distribution are also uniformly distributed.
When the mean of the distribution is defined by the weights we use in production and we randomise purely for the sake of data collection, this seems logical and desirable.

Nevertheless, when we consider higher dimensions, this property vanishes.
Imagine we sample 1 million points from a uniform distribution in $\mathcal{U}(-0.5, 0.5)$.
In a single dimension, it is easily verified that the (normalised) Euclidean distance of every point is also uniformly distributed.\footnote{Because the maximal Euclidean distance depends on the dimensionality $d$, we normalise the distances by a factor $\sqrt{0.25d}$ to allow comparison across changing $d$.}
If we increase the number of dimensions, however, this desirable property no longer holds.
Suppose we were to sample from a Gaussian distribution instead: this bias is also present, but in a far less pronounced manner.

Intuitively, what practitioners \emph{expect} when choosing a uniformly distributed data collection policy is that the probability of a point falling into a region $[\mu-\epsilon,\mu+\epsilon]$ (where $\mu$ is the centre of the distribution) grows linearly with $\epsilon$ (until the upper bound of the interval).
This is clearly valid for the univariate case.
Nevertheless, this is \emph{not} what we get for the multivariate case.
Indeed, this probability can be computed as $\mathsf{P}(x \in [\mu-\epsilon,\mu+\epsilon]) = (1-2\epsilon)^{d}$, implying that a hypersphere with a radius covering 80\% of the maximal distance to the origin ($\epsilon=0.4$), only covers roughly 16\% of the data points for $d=8$.
This fraction decreases monotonically as $d$ grows.

We visualise these phenomena in Figure~\ref{fig:cod}, where we sample 1 million points from a Uniform distribution, and 1 million points from an isotropic Normal distribution.
The left and middle plot visualise the cumulative density of the normalised Euclidean distance to the mean of the distribution (i.e. the origin), over sampled points.
The linear line on the left-hand plot (i.e. $\mathcal{U} (d=1)$) exhibits the desirable uniform-distance property, which is not retained for increasing $d$.
We plot the empirical fraction of sampled points falling into a centred hypersphere with radius $\epsilon$ on the right-hand plot.
Analogous to the cumulative density, whilst we observe a linear trend for $\mathcal{U} (d=1)$ this quickly changes for increasing $d$.
The multivariate normal distribution provides a smoother transition, with 75\% of points being covered for $\epsilon=0.4$ at $d=8$, instead of 16\%.

A potential downside of using a multivariate normal logging distribution instead of a uniform distribution, is that its domain is unbounded.
Nevertheless, the cumulative density plots show that the probability of observing extreme points is negligible for practical purposes.
We gain the desirable property of more sampled points falling near the mean of the distribution, which in turn implies a higher effective sample size when we wish to evaluate policies that make small, rather than drastic changes to the production weights.

Furthermore, common support for the logging and target policies opens the door to many statistical tools and tricks that allow for more effective and efficient estimation for both evaluation and learning: baseline corrections~\cite{Gupta2024_baselines}, and self-normalisation~\cite{Swaminathan2015}.

\section{Designing a sensitive reward signal}\label{sec:learntreward}
The algorithmic decision-making lens that motivates the use of policy learning techniques for recommendation use-cases, allows us to directly target key online metrics as the \emph{reward} signal for the learnt policy.
Whilst \emph{effective}, typical North Star metrics such as long-term growth might not be \emph{efficient} labels to consider.
Indeed, the variance of the estimators discussed in Section~\ref{sec:crm} depends on the inherent variance of the reward signal in the estimand.
As discussed in Section~\ref{sec:ESS}, the effective sample size is also dependent on this reward, further highlighting its importance as a crucial design consideration in the modelling process.

In essence, we wish to use a reward signal that is highly (causally) correlated with the North Star, but potentially has lower variance.
Variance reduction techniques are a staple in the online experimentation research literature, typically leveraging regression adjustments as additive control variates~\cite{Deng2013, Poyarkov2016, Budylin2018, Baweja2024}.
Note that this line of work is analogous to the use of Doubly Robust estimators in the off-policy learning literature~\cite{Dudik2014,Jeunen2020REVEAL}.
As these techniques would require us to introduce an additional model that estimates the reward, they are out-of-scope for the purposes of this work.

Other work has focused on ``\emph{data-driven metric development}''~\cite{Deng2016}, \emph{learning} online metrics that directly target statistical sensitivity~\cite{Kharitonov2017}.
We leverage the work of \citet{Jeunen2024_Learning} to learn a reward signal that minimises a convex lower bound on the number of type-III and type-II errors the metric exhibits on a logged dataset of past A/B-experiments.
This minimises observed $p$-values and, hence, maximises statistical power w.r.t. the North Star.
We refer the interested reader to their work for technical details~\cite{Jeunen2024_Learning}, and evaluate the sensitivity of this learnt metric as a reward signal for off-policy learning and evaluation in Section~\ref{sec:experiments}.
\section{Experiments \& Discussion}\label{sec:experiments}
In this section, we wish to empirically validate the theoretical and technical contributions set forth earlier in this article.
The research questions we wish to answer can be summarised as:
\begin{description}
    \item[\textbf{RQ1}] \textit{Can effective sample size corrections improve the coverage of estimator confidence intervals for (SN)IPS?}
    \item[\textbf{RQ2}] \textit{Does a learnt reward signal affect the sensitivity of results?}
    \item[\textbf{RQ3}] \textit{Does our proposed approach lead to offline improvements?}
    \item[\textbf{RQ4}] \textit{Does our proposed approach lead to online improvements?}
\end{description}

All source code to reproduce the empirical results on synthetic data can be found at \href{https://github.com/olivierjeunen/multivariate-policy-learning-recsys-2024}{github.com/olivierjeunen/multivariate-policy-learning-recsys-2024}.

\begin{figure*}[!t]
    \centering
    \vspace{-3ex}
    \includegraphics[width=\textwidth]{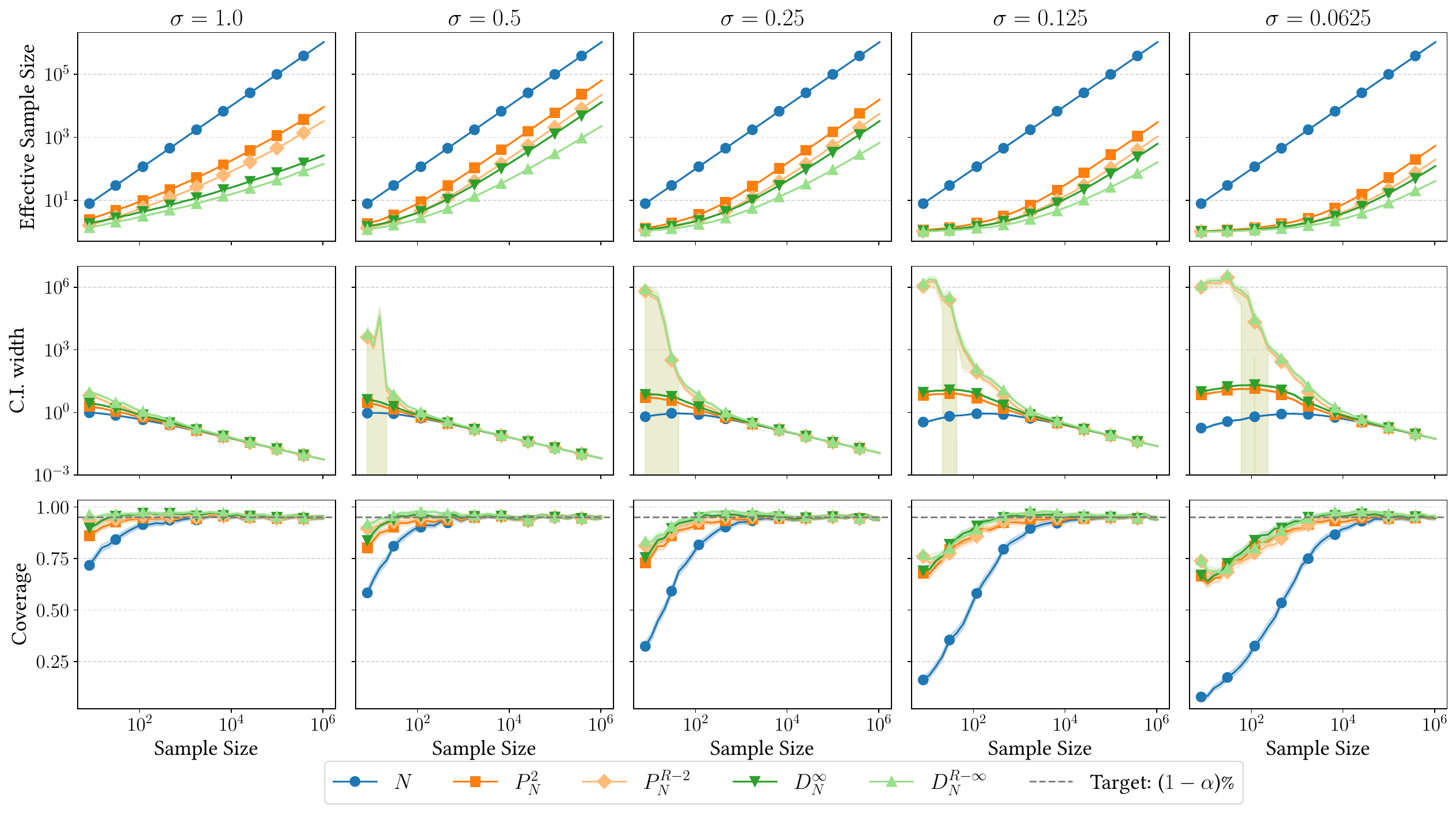}
    \caption{Off-policy evaluation results for $\widehat{V}_{\rm SNIPS}$, considering the coverage of C.I.s obtained through various variations of our proposed ESS correction. ESS-corrected methods attain target coverage at significantly reduced sample sizes.}
    \label{fig:coverage}
\end{figure*}
\begin{figure}[!t]
    \centering
    \vspace{-3ex}
    \includegraphics[width=0.9\linewidth]{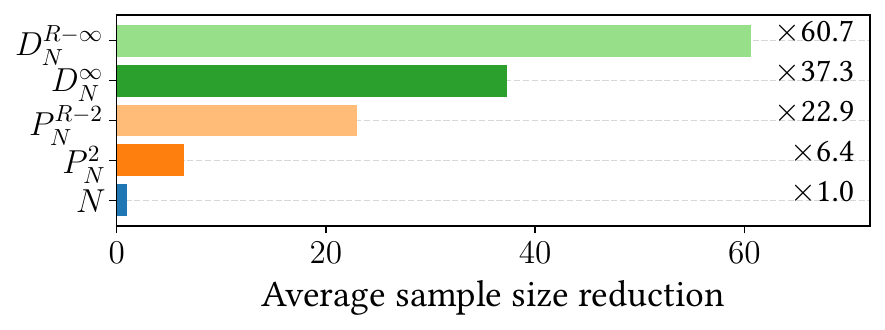}
    \caption{All considered C.I. corrections significantly reduce the required sample size for the C.I. to achieve its specified coverage level: down to 60$\times$ on average for $D_{N}^{R-\infty}$.}
    \label{fig:agg_reduction}
\end{figure}

\subsection{ESS Corrections \& Estimator Coverage (RQ1)}\label{sec:synthetic}
As argued in Section~\ref{sec:ESS}: whilst existing variance estimators for the CRM objective are guaranteed to hold in the limit of large sample sizes because of the CLT, they might give rise to confidence interval that exhibit insufficient coverage in finite sample scenarios.
We wish to validate whether the policy-dependent correction laid out in Section~\ref{sec:ESS} can improve empirical coverage, when using a range of potential estimators for the effective sample size.

Naturally, the statistical coverage of a confidence interval can only be validated via simulation studies using synthetic data.
This has the added advantage of ease-of-reproducibility.

We simulate a multivariate isotropic Gaussian logging policy with identity co-variance and $d=5$ dimensions, centred at the origin.
We shift the target policies to be centred at $\mu_{t}=0.5$, and consider varying standard deviations $\sigma \in \{1, 0.5, 0.25, 0.125, 0.0625\}$.
To simulate a realistic setting where the target is e.g. the number of active days a user spends on the platform, we sample Poisson-distributed rewards where the rate is proportional to the average of the weights $\bar{a}$: $r \sim {\rm Poisson}(\max(0,0.1\cdot\bar{a}))$.
This allows us to easily recover the ground truth policy value, as $\mathbb{E}_{a \sim \mathcal{N}(\mu_t,\sigma I)}[R] = 0.1\cdot\mu_t$.
We sample between $2^{3}$ and $2^{20}$ data points from $\pi_{0}$ and use these to estimate the value of the target policy $\widehat{V}_{\rm (SN)IPS}(\pi_{t})$.
We repeat this process $2\,000$ times to smooth out the effects of stochasticity.

Confidence Intervals (C.I.s) for $\widehat{V}_{\rm (SN)IPS}(\pi_{t})$ are obtained via Eq.~\ref{eq:better_LB}, with varying estimators for $\widehat{{\rm ESS}}$ and thus $\widetilde{N}$.
We report three key metrics:
\begin{enumerate*}
    \item the estimated ESS (directly proportional to the discrepancy between $\pi_{0}$ and $\pi_{t}$,
    \item the width of the C.I. (directly proportional to the estimated variance of the mean), and
    \item the coverage of the 95\% C.I. (which should converge to 95\%).
\end{enumerate*}
We increase the sample size over the $x$-axis, and show results for varying target policy standard deviations in different columns.
Results for $\widehat{V}_{\rm SNIPS}$ are visualised in Figure~\ref{fig:coverage}.

Key observations from this plot include that the traditional sample variance C.I. has poor coverage in realistic scenarios, when sample sizes are relatively low (up to 10k in this simple non-contextual setting).
Convergence does indeed occur, as guaranteed by the CLT, but slowly. Furthermore, it is unclear \emph{when} we can expect the CLT to hold.
In constrast, all the proposed ESS estimators combined with our sample size correction improve coverage at smaller sample sizes, whilst still converging to the advertised coverage level as the sample size grows.
These empirical findings corroborate our theoretical expectations.
Pessimistic ESS estimates, i.e. those leveraging the $\ell_{\infty}$-norm proposed by \citeauthor{Martino2017}~\cite{Martino2017}, outperform the conventional estimates that rely on $\ell_{2}$-norms.
Reward-dependent ESS estimates, as discussed by \citeauthor{Owen2013}~\cite{Owen2013}, outperform their reward-independent counterparts.
The most promising variant is $D_{N}^{R-\infty}$, the estimator that combines both pre-existing elements in the literature into a novel estimator for the ESS.
Importantly, we note that the corrections do not significantly overshoot the required coverage level, and converge to the CLT C.I.s as $N$ grows.

Aggregating for all considered target policies the ratio of the minimum sample size required to achieve the advertised coverage level using $N$ and using $\widetilde{N}$, we observe average sample size reductions of up to 60 times (i.e. down to 1.7\% of the traditional C.I.).
These results are visualised in Figure~\ref{fig:agg_reduction}.
This represents a significant shift that improves the robustness of offline evaluation and learning capabilities for multivariate policy learning models in production.
Results for $\widehat{V}_{\rm IPS}$ are qualitatively similar but omitted for brevity.

\begin{figure*}[!t]
\centering
\vspace{-3ex}
\begin{subfigure}[t]{0.475\linewidth}
    \centering
    \includegraphics[width=\linewidth]{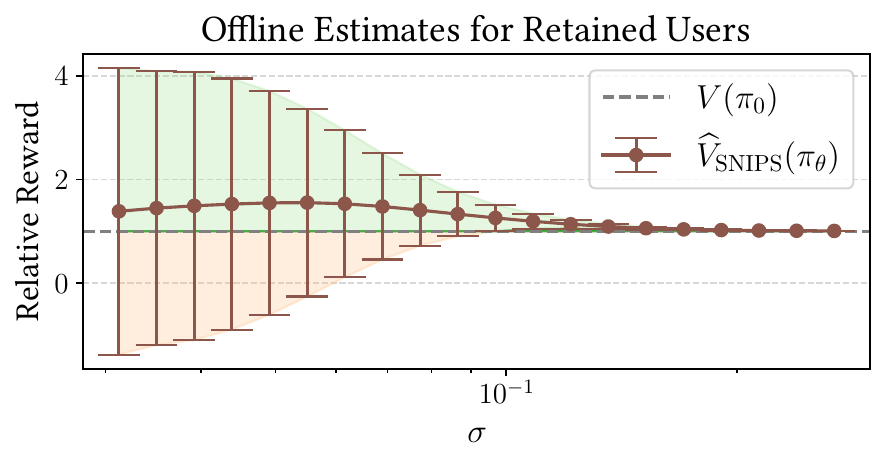}
    \vspace{-5ex}
    \caption{Off-policy evaluation results for the ``retained users'' label. High variance leads to inconclusive results and extremely wide C.I.s.}
\end{subfigure}\hfil
\begin{subfigure}[t]{0.475\linewidth}
    \centering
    \includegraphics[width=\linewidth]{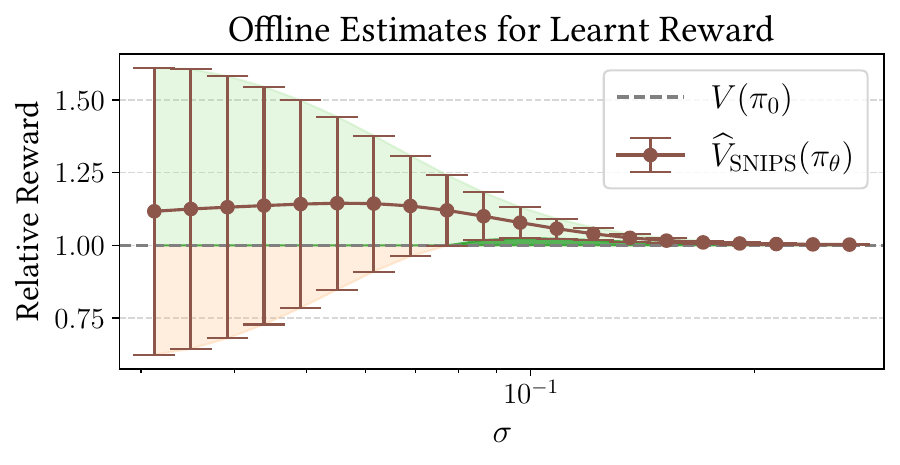}
    \vspace{-5ex}
    \caption{Off-policy evaluation results for the ``learnt reward'' label. Around $\sigma\approx0.1$, improvements are statistically significant with $p<0.05$.}
 \end{subfigure}
 \caption{Off-policy evaluation results for varying $\sigma$, visualising the bias-variance trade-off that comes with the kernel smoothing technique. We provide results for an insensitive North Star (a), and a learnt reward signal that maximises statistical power (b).}\label{fig:offline}
\end{figure*}

\subsection{Improved Recommendations (RQ2--4)}
To empirically validate that we can leverage the multivariate policy learning approach to learn improved policies and ultimately improve the recommendations that are being shown to end-users, we require either access to
\begin{enumerate*}
    \item a dataset where weights were actively randomised and logged along with the appropriate reward signals, or
    \item a simulation environment that accurately mimics real user behaviour in response to recommendations, over multiple signals and objectives (e.g. engagements, time spent, retention).
\end{enumerate*}
Whilst the latter has been popular to validate general off-policy learning approaches to recommendation~\cite{Sakhi2020,Jeunen2020,Jeunen2021_Pessimism,Jeunen2023,Liu2022}, existing simulators only consider single objectives (e.g. clicks) and it is unclear how these results translate to experiments with real end-users.
Because no appropriate datasets are publicly available, we resort to proprietary datasets obtained from large-scale short-video platforms.
We report both offline evaluation results and results from online experiments in the following subsections.

The base recommendation model on the platforms is a Multi-gate Mixture of Experts (MMOE) multi-task learning model~\cite{Ma2018}, used to generate predictions for 9 different behaviour signals (i.e. like, share, favourite, end session and view, among others).
These are then combined following Eq.~\ref{eq:scalarisation}.

\subsubsection{Production Baseline (Control)}
The production baseline can be seen as an instantiation of the Direct Method~\cite{Dudik2014}, where a direct search is performed to find the scalarisation weights that maximise the correlation with retention, typically seen as a strong indicator of user satisfaction and long-term growth for the platform.
These deterministic production weights are then treated as hyper-parameters of the overall system, and have undergone multiple manual tuning rounds where the outcomes of online experiments decide which weights will be used going forward.
Because online performance is measured directly, this approach is effective.
Nevertheless, because the search space of possible weights is vast, the approach is inefficient and costly.
Automated policy learning methods that can improve over the production weights are therefore of significant importance to the platform.

\subsubsection{Offline Results (RQ2--3)}\label{sec:offline}
To obtain an offline dataset to learn from, we randomise weights for 4.5 million users and log their behaviour on the platform over a period of three weeks.
The randomised weights correspond to the \emph{actions}, and the behaviours inform the \emph{rewards} we wish to optimise using our proposed multivariate off-policy learning approach.

We optimise a deterministic policy to maximise a lower bound on the $\widehat{V}_{\rm SNIPS}$ estimator, given by Eq.~\ref{eq:better_LB}.
For the learnt policy, we visualise the obtained ESS-corrected confidence interval for the value estimates when we vary the bandwidth of the Gaussian kernel (i.e. the standard deviation $\sigma$).
These results are visualised in Figure~\ref{fig:offline}.
The policy we wish to deploy is deterministic, i.e. $\sigma\equiv 0$.
Naturally, this would lead to an extremely low ESS, and high variance as a result.
As we increase $\sigma$, we observe that the variance of the estimator \emph{decreases} as its bias \emph{increases}.
Indeed, for large values of $\sigma$, the estimates coincide with the empirical reward obtained by the logging policy, as the widening Gaussian kernel tends to smooth over \emph{all} observed samples. 
This visualises the bias--variance trade-off that occurs for off-policy evaluation in continuous action domains.

We plot offline evaluation results for two reward signals: ``retained users'' as a direct but noisy measurement of platform growth objectives, and our ``learnt reward'' signal introduced in Section~\ref{sec:learntreward}.
First, we observe that offline estimates for the ``retained users'' reward are directionally positive, but imply such high variance that we cannot make confident statements about the superiority of the learnt policy.
Furthermore, this inhibits any meaningful estimates about the treatment effect of the learnt policy, as the C.I.s indicate an increase of up to 300\%, which no recommendation policy change could ever reasonably incur.

When considering the learnt reward signal instead, we first observe a far more reasonable scale on the y-axis.
Second, we observe that around $\sigma\approx0.1$ the 95\% confidence intervals denoting relative improvements over the logging policy no longer include 1.
In other words, the improvements are statistically significant with $p<0.05$. 
This leads us to consider the learnt policy as a promising candidate for an online experiment.

\begin{table*}[!t]
\begin{subtable}[t]{0.45\textwidth}

    \centering
    \begin{tabular}{lc}
    \toprule
        \textbf{Metric} & \textbf{Relative Improvement (95\% C.I.)} \\
    \midrule
    Retention & \colorbox{gray!10}{[--0.05\%, +0.07\%]} \\
    Time on Platform  & \colorbox{gray!10}{[--0.07\%, +0.34\%]} \\
    Heavy Users & \colorbox{LimeGreen!50}{[+0.07\%, +0.30\%]} \\
    Learnt Reward & \colorbox{LimeGreen!50}{[+0.04\%, +0.27\%]}  \\
    \bottomrule
    \end{tabular}
    \caption{Global weights on platform A, 14 days, 6.4 million users.}
    \label{tab:AB1}
\end{subtable}
\hspace{0.05\textwidth}
~\hfill~
\begin{subtable}[t]{0.45\textwidth}
    \centering
    \begin{tabular}{lc}
    \toprule
        \textbf{Metric} & \textbf{Relative Improvement (95\% C.I.)} \\
    \midrule
    Retention & \colorbox{LimeGreen!50}{[+0.01\%, +0.25\%]} \\
    Time on Platform & \colorbox{LimeGreen!50}{[+0.08\%, +1.14\%]} \\
    Heavy Users & \colorbox{gray!10}{[--0.07\%, +0.51\%]} \\
    Learnt Reward & \colorbox{LimeGreen!50}{[+0.03\%, +0.51\%]}  \\
    \bottomrule
    \end{tabular}
    \caption{Feed-specific weights on platform B, 14 days, 10 million users.}
    \label{tab:AB2}
\end{subtable}
\caption{Online A/B-test results from deploying our approach on two large-scale short-video platforms.
We report Bonferroni-corrected 95\% confidence intervals for key platform metrics.
We observe statistically significant improvements to both the optimisation target and adjacent metrics, \colorbox{LimeGreen!50}{highlighting} results that are statistically significant with $p<0.05$.}
\end{table*}

\subsubsection{Online Results (RQ4)}
We deploy the learnt policy obtained through the offline experiment in Figure~\ref{fig:offline} in an online A/B-test with 6.4 million users, over the course of 2 weeks.
Results are reported in Table~\ref{tab:AB1}, for various metrics that are key to the platform.
\textbf{Retention} denotes the fraction of users that use the app the day after, \textbf{Time} on the platform denotes overall time spent on the app, \textbf{Heavy Users} are users that have more than $X$ positive item interactions, and the \textbf{Learnt Reward} is a metric specifically optimised to maximise statistical power w.r.t. the North Star, which is also the optimisation target for our policy learning approach, as described in Section~\ref{sec:learntreward}.
We apply a Bonferroni-correction to obtain 95\% C.I.s for the relative improvement in these metrics, for the learnt weights over the production weights.
We observe that the learnt policy is able to improve the target metric in the online test, a testament to the effectiveness of the off-policy learning framework.
Indeed, mismatches between online and offline evaluation results plague the recommender systems research field~\cite{Jeunen2019,Gruson2019,Gilotte2018}, but the decision-making lens allows us to directly optimise offline estimators of online metrics instead~\cite{Jeunen2021_Thesis}.
This experiment highlights the effectiveness of such approaches for general multi-objective recommendation scenarios.

To reproduce, validate and extend these empirical insights, we perform a second online A/B-test on another large-scale short-video platform, considering similar metrics.
We report results in Table~\ref{tab:AB2}.
In this experiment, we learn different scalarisation weights for different short-video feeds users can interact with on the platform (i.e. home page feed, specific feeds, et cetera).
Reassuringly, we also observe that the learnt policy is able to improve the target metric significantly.
The time users spend on the platform and user retention also improve significantly.
Because these metrics are not easily moved by online experiments, this represents an important improvement to end-users' experience on the platforms, and signifies the effectiveness of our method.

\section{Conclusions \& Outlook}
Recommender Systems operating on online platforms are typically optimised for multiple objectives via scalarisation techniques~\cite{Hwang1979}.
Whilst these techniques help to find Pareto-optimal solutions, the choice of \emph{where} on the Pareto-front the platform wishes to place itself, remains a question that is hard to answer.

In this work, we present a general approach to solve this problem, leveraging elements from the algorithmic decision-making literature.
In particular, we propose to use the Counterfactual Risk Minimisation paradigm~\cite{Swaminathan2015_BLBF} with a continuous multivariate action space to learn a scalarisation policy that maximises a notion of North Star reward defined by the platform.
To make our approach more effective and efficient, we propose a policy-dependent correction on the lower bound that is typically used by CRM, which improves empirical coverage at small sample sizes whilst retaining CLT guarantees in the limit.
In doing so, we provide a novel estimator for the ESS and empirically show its utility using synthetic data.
Furthermore, we highlight that the common practice of preferring \emph{uniform} randomisation is counterproductive in multivariate domains, and propose to leverage work on ``learnt metrics'' to design reward signals that are highly correlated with the North Star, but bring improved statistical power to benefit our learning method~\cite{Jeunen2024_Learning}.

We apply our learning method to two large-scale short-video platforms, with over 160 million monthly active users each, and showcase via off- and online experiments that our methods are able to bring significant value to the platforms' objectives.

\bibliographystyle{ACM-Reference-Format}
\bibliography{bibliography}

\end{document}